\documentclass[preprints,article,accept,moreauthors,pdftex]{mdpi} 

\firstpage{1} 
\pubvolume{1}
\issuenum{1}
\articlenumber{0}
\pubyear{2021}
\copyrightyear{2021}
\datereceived{7/29/21} 
\hreflink{https://doi.org/} 
\pdfoutput=1

\Title{Neural Upscaling from Residue-level Protein Structure Networks to Atomistic Structures}

\TitleCitation{Neural Upscaling from Residue-level Protein Structure Networks to Atomistic Structures}

\Author{Vy T. Duong $^{1}$, Elizabeth M. Diessner $^{1}$, Gianmarc Grazioli $^{2}$, Rachel W. Martin $^{1,3^{\ast}}$, Carter T. Butts $^{4^{\ast}}$
}

\AuthorNames{Vy Duong, Elizabeth Diessner, Gianmarc Grazioli, Rachel W. Martin, Carter T. Butts}

\AuthorCitation{Duong, V; Diessner, E.; Grazioli, G; Martin, R.W.; Butts, C.T.}

\address{%
$^{1}$ \quad Department of Chemistry, UC Irvine\\
$^{2}$ \quad Department of Chemistry, San Jose State University\\
$^{3}$ \quad Department of Molecular Biology \& Biochemistry, UC Irvine\\
$^{4}$ \quad Departments of Sociology, Statistics, and Electrical Engineering \& Computer Science, UC Irvine
}

\corres{Correspondence: buttsc@uci.edu,rwmartin@uci.edu;}

\abstract{
Coarse-graining is a powerful tool for extending the reach of dynamic models of proteins and other biological macromolecules.  Topological coarse-graining, in which biomolecules or sets thereof are represented via graph structures, is a particularly useful way of obtaining highly compressed representations of molecular structure, and simulations operating via such representations can achieve substantial computational savings.  A drawback of coarse-graining, however, is the loss of atomistic detail - an effect that is especially acute for topological representations such as protein structure networks (PSNs).  Here, we introduce an approach based on a combination of machine learning and physically-guided refinement for inferring atomic coordinates from PSNs.  This ``neural upscaling'' procedure exploits the constraints implied by PSNs on possible configurations, as well as differences in the likelihood of observing different configurations with the same PSN.  Using a 1 $\mu$s atomistic molecular dynamics trajectory of A$\beta_{1-40}$, we show that neural upscaling is able to effectively recapitulate detailed structural information for intrinsically disordered proteins, being particularly successful in recovering features such as transient secondary structure.  These results suggest that scalable network-based models for protein structure and dynamics may be used in settings where atomistic detail is desired, with upscaling employed to impute atomic coordinates from PSNs.}


\begin{document}

\section{Background}
Proteins and other biological macromolecules exhibit a wide variety of complex dynamics and interactions at varying size and time scales. While atomistic molecular dynamics (MD) models currently serve as the gold standard tools for simulating dynamics at high resolution (with some inroads by quantum mechanical methods in small-scale  or specialized applications), the cost of large-scale MD simulations limits their use to relatively small systems on time scales of microseconds or less.  Coarse-grained (CG) models offer a means of accessing larger system sizes and longer time scales, sacrificing atomistic detail in exchange for reduced computational cost.  Many ``flavors'' of coarse-grained simulation exist, with the most common being aggregate particle models that represent collections of atoms by single particles whose positions evolve under a suitably modified forcefield.  The highly successful MARTINI model \cite{marrink2007martini}, for instance, represents biomolecules by ``beads'' corresponding roughly to one bead per four heavy atoms, with hydrogens left implicit; MARTINI and other CG MD models have proven useful in studying the structure and dynamics of large complexes, lipid phases, and other systems that are too large to be treated with atomistic MD methods \cite{capelli_data-driven_2021}.  An even more radical approach to coarse-graining employs \emph{topological} representations, representing molecules or molecular aggregates by network structures that encode the interactions between atoms or groups thereof, but not their positions in three-dimensional space \cite{benson.daggett:jbcb:2012,mustoe_coarse_2014}.  Although most often employed for descriptive analysis of trajectories produced by MD or other methods (see e.g. \cite{wong.et.al:bio:2019,cross.et.al:bio:2020,demakis_conserved_2021}), recent work has also considered the generation of trajectories directly within the topological representation, allowing considerable computational savings \cite{grazioli_comparative_2019,grazioli2019network}. 

While many questions can be posed directly within a CG representation, an obvious limitation of coarse-graining is that some observables of interest cannot be obtained without an additional step of ``backmapping'' or ``upscaling'' the CG trajectory to atomistic resolution.  At first blush, this may seem impossible: by definition, a CG model does not resolve individual atoms.  In practice, however, CG structures are often sufficiently constraining that a well-designed algorithm can infer atomic positions from them with considerable accuracy \cite{ferrie_unified_2020}.  For instance, a number of upscaling methods for particle-based CG models work via a two-stage process in which initial guesses for atomic placement are made based on e.g. random positioning \cite{rzepiela2010reconstruction}, fragment-based \cite{hess2006long,peter2009multiscale}, or geometry-based \cite{gopal2010primo,brocos2012multiscale,wassenaar2014going,machado2016sirah} initialization, followed by an energy minimization step to ensure physically realistic coordinates.  This is not unrelated to protein structure prediction methods like those of \cite{bonneau_rosetta_2001,zhang_template-based_2007}, which begin with approximate structures based on local homology and subsequently refine them via minimization in a simplified force field.  Such techniques have proven extremely successful in predicting the structure of globular proteins \cite{tyka_alternate_2011,pearce_toward_2021}, and are widely used in enzyme discovery and engineering applications \cite{smith_assessing_2020,alford_rosetta_2017}.

In the context of topological coarse-graining, the historical focus has been on mapping from atomistic to coarse-grain networks for purpose of analysis (e.g. \cite{benson.daggett:jbcb:2012,webb2018graph,chakraborty2018encoding,unhelkar.et.al:bba:2017,duong.et.al:ib:2018}), with correspondingly less emphasis on the upscaling problem.  Recent work, however, has suggested the potential of graph-theoretic models for molecular structure and dynamics.  For instance, \citet{grazioli2019network,yu.et.al:nsr:2020} use Hamiltonians defined on graphs representing the structures of protein aggregates to model the equilibrium structures and kinetics of amyloid fibrils and associated aggregation states (with vertices representing individual proteins, and edges indicating bound interactions).  On a smaller scale, \citet{grazioli_comparative_2019} used a closely related approach to model transient structure in intrinsically disordered proteins (IDPs), using residue-level protein structure networks (PSNs) in which each vertex represents a residue and edges represent inter-residue contacts.  Although we are unaware of any existing methods for upscaling such graph structures to atomic resolution, effective methods for this purpose would greatly extend the practical reach of network-based simulation models. 

\begin{figure}[H]
	\centering
	\includegraphics[width=0.7\textwidth]{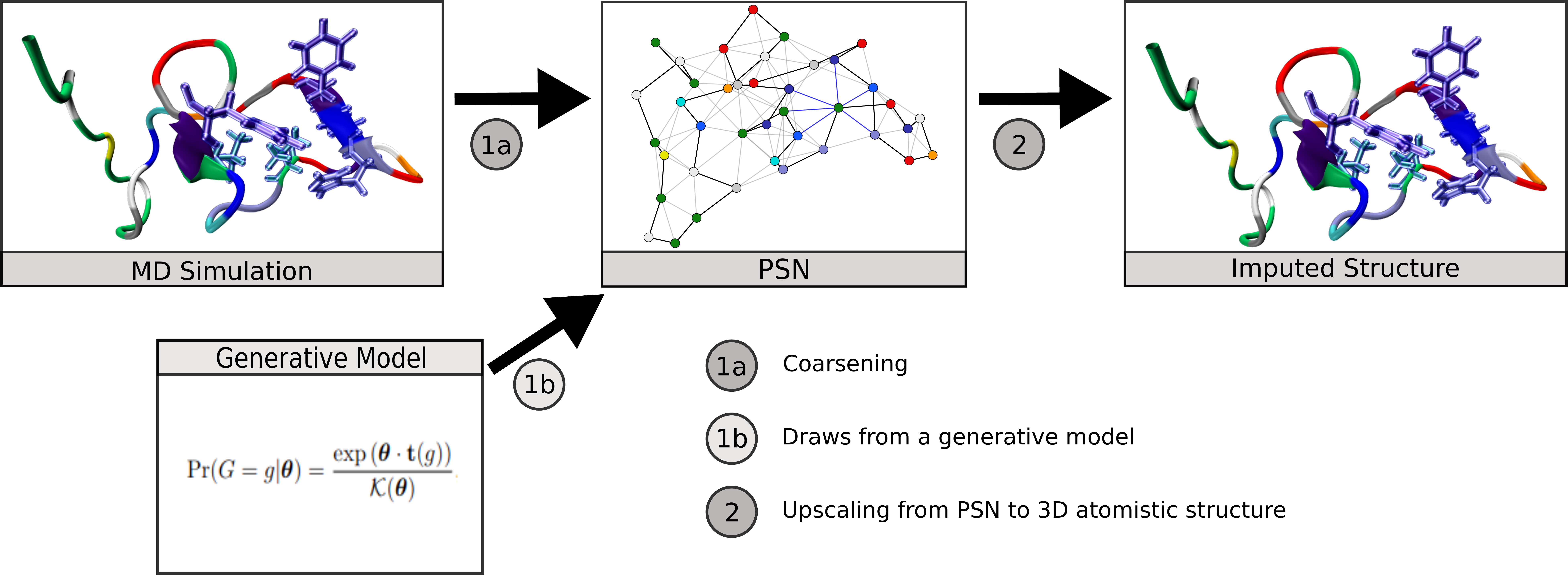}
	\caption[The ability to impute atomistic structures from network representations]{The ability to impute atomistic structures from network representations enables both compressive representation of structures from MD trajectories and the use of generative network models to predict distributions of atomistic structures.}
	\label{NN_supp_imputed}
\end{figure}

Our focus in this paper is this last problem: the upscaling of topological representations of macromolecular structure (and by extension, dynamics) to atomic resolution.  We specifically consider the upscaling of residue-level PSNs, as this is a widely used level of network coarse-graining for proteins and poses a non-trivial challenge for atomistic refinement.  To perform the mapping from network structure to atomistic structure, we exploit advances in machine learning (ML) methods, predicting atomic coordinates from topological inputs using deep neural networks.  Machine learning strategies (particularly including neural networks) have become widely used in CG modeling, with past efforts focused on ML-based methods for learning or refining CG forcefields (see e.g.  \cite{bejagam2018machine,boninsegna2015investigating,chakraborty2018encoding,lemke2017neural,wang2019machine,webb2018graph,zhang2018deepcg}). Here, we use multilayer perceptron-based (MLP) neural networks to learn pairwise interatomic distances from residue-level PSNs, allowing us to recover atomistic detail from input network structures.

In this work, we demonstrate the utility of MLP neural network models to translate coarse-grained protein structure network representations to their more finely detailed 3D coordinate structures. We apply this to the case of IDPs, showing that the trained neural network is able reproduce equilibrium conformations of amyloid-$\beta$ protein obtained from MD simulations at atomic-level detail, also capturing its diverse transient secondary structure behavior.  We additionally consider the use of further refinements (such as chirality corrections and energy minimization) to improve predictive performance.  We show that this scheme can obtain a high level of accuracy, with median RMSE for predicted versus true 3D structures of approximately 2.13 \AA{ } and a high degree of correspondence for relatively folded regions of the protein.  The resulting scheme provides a practical mechanism for mapping PSNs produced by generative network models to predicted atomistic structures (Figure \ref{NN_supp_imputed}), for using PSNs as an efficient tool for lossy compression of long trajectories, or other applications in which it is useful to infer atomistic information from coarse-grained topological representations.

The remainder of the paper is organized as follows.  We introduce our approach in Sec.~\ref{sec_methods}, including both our ML pipeline and subsequent refinement methods.  Sec.~\ref{sec_results} reports the results of our simulation study, Sec.~\ref{sec_disc} discusses further directions, and Sec.~\ref{sec_conclusion} concludes the paper.


\section{Methods} \label{sec_methods}

\begin{figure}[H]
	\centering
	\includegraphics[width=0.5\textwidth]{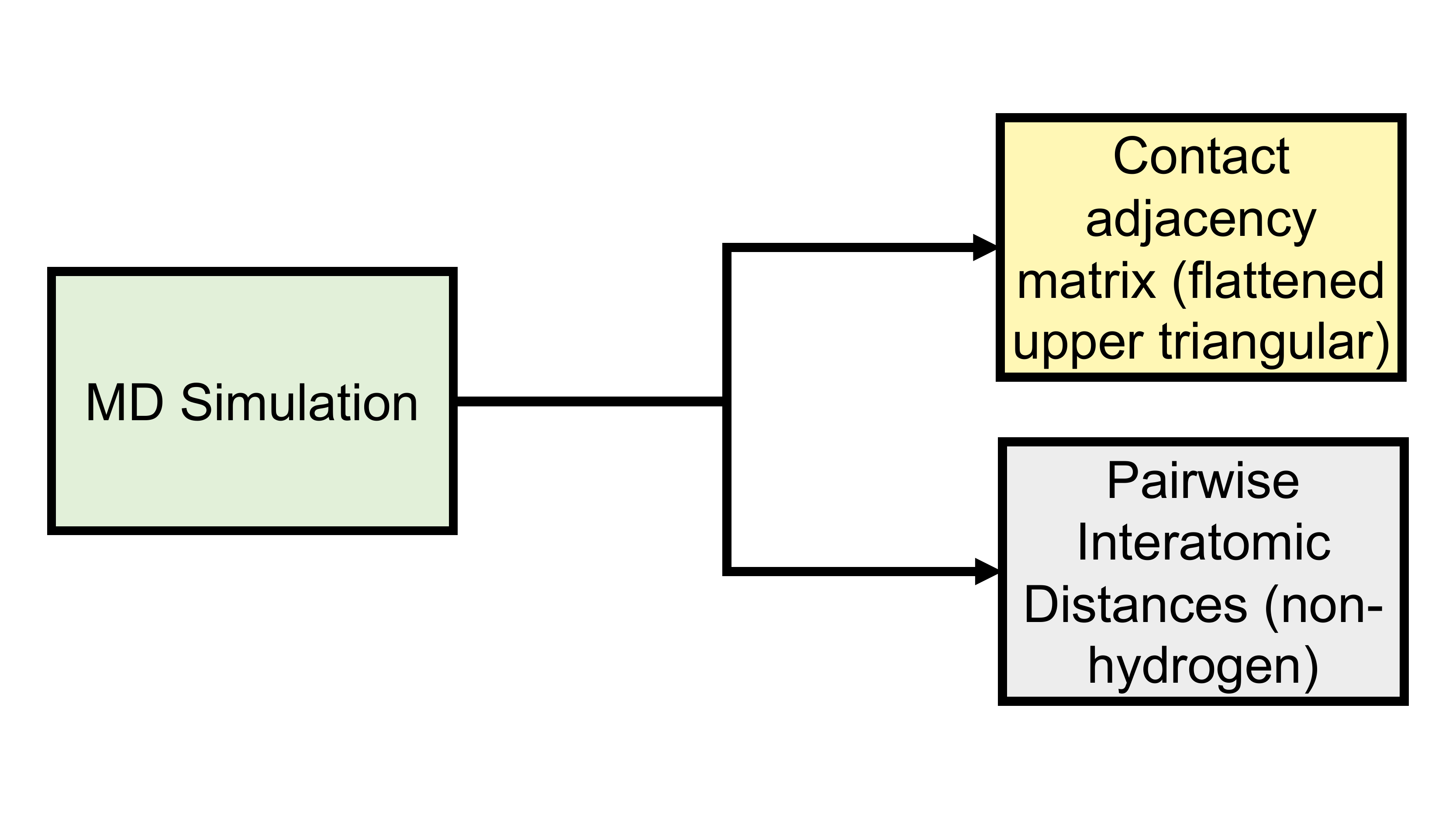}
	\caption[Data generation of input and output data.]{Data generation of input (upper triangle of PSN adjacency matrices) and output (upper triangular of PIDs) data. }
	\label{NN_method}
\end{figure}

\textbf{Data Generation} Our data come from atomistic MD trajectories of A$\beta_{1-40}$, a widely studied IDP implicated in the etiology of Alzheimer's disease; the atomistic trajectories and associated PSN coarsenings respectively serve as ground truth and inputs for the upscaling model (Figure~\ref{NN_method}). Beginning with the lowest energy monomer of the PDB structure, 2LFM \cite{vivekanandan.et.al:bbrc:2011}, one A$\beta_{1-40}$ monomer was simulated in explicit solvent for 1 $\mu$s using NAMD \cite{phillips.et.al:jcc:2005} via the following protocol: the initial monomer structure was solvated in a cubic TIP3P  \cite{jorgensen.et.al:jcp:1983} water box of minimum margin 25 Angstroms, and neutralized with NaCl counter-ions. This assembly was minimized for 10,000 iterations, followed by velocity initialization and 250 simulation iterations before final adjustment of the water box.  A trajectory of approximately $1.1 \mu$s was then simulated. Simulation was performed under periodic boundary conditions in NAMD with the CHARMM36m forcefield \cite{huang.et.al:nm:2017}, using an NPT ensemble at 300K and 1 atm pressure. Temperature control was maintained by Langevin dynamics with a period of 1/ps, with Nos\'e-Hoover Langevin piston pressure control \cite{martyna.et.al:jcp:1994,feller.et.al:jcp:1995}.  Creation of initial conditions and related data processing were performed using VMD \cite{humphrey.et.al:jmg:1996}.

The simulation contains 11,926 total frames/conformations, of which 72\% was allocated for training, 20\% for testing, and 8\% for validation. 5-fold cross validation was also performed to ensure bias was not introduced during initial train-test splitting. For each frame in the A$\beta_{1-40}$ simulation, a protein structure network (PSN) was calculated using a combination of VMD \cite{humphrey.et.al:jmg:1996} and the \texttt{statnet} \cite{handcock.et.al:jss:2008,butts:jss:2008a} and \texttt{bio3d} \cite{grant.et.al:b:2006} libraries for \textsf{R} \cite{rteam:sw:2018}). 

Monomer states were sampled from the trajectory every 100 ps, from which residue-level protein structure networks were constructed. Vertices correspond to individual residues, with two vertices being considered adjacent if they contain respective atoms whose distance is less than or equal to 1.1 times the sum of their van der Waals radii (based on radius data from \cite{alvarez:dt:2013}). The input data used to train the neural network model consists of the flattened upper triangular matrix data extracted from the residue-level contact adjacency matrix for each conformation in the A$\beta_[1-40]$ trajectory. The output data used to train the model is the flattened upper triangular of pairwise interatomic distance matrices (PIDs) calculated for each non-hydrogen atom (across all frames in the MD simulation) (Figure \ref{NN_method}).


\begin{figure}[H]
	\centering
	\includegraphics[width=0.7\textwidth]{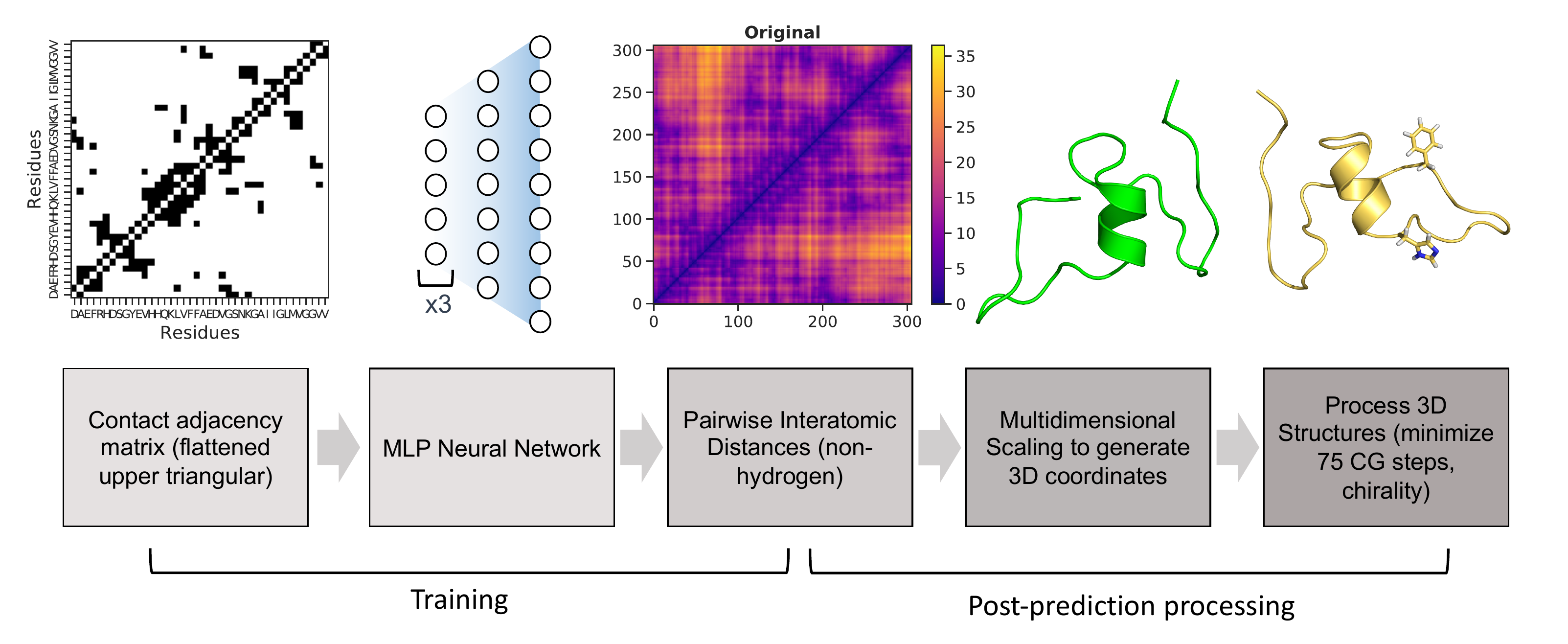}
	\caption[Pipeline of MLP neural network training and post-prediction processing.]{Pipeline of MLP neural network training and post-prediction processing.}
	\label{NN_method2}
\end{figure}

\textbf{Neural network architecture and hyperparameters} After generation of input and output data, a multi-layer perceptron (MLP) neural network was utilized for training as indicated in the pipeline (Figure \ref{NN_method2}). The neural network is based on a multi-layer perceptron utilizing the machine-learning libraries Keras \cite{chollet2015keras} and tensorflow \cite{abadi2016tensorflow}. The first three hidden layers consist of 2000 neurons, the fourth layer contains 8000 neurons, and the last output layer predicts the flattened upper triangular of the pairwise interatomic distance matrix for a given frame from the MD simulation (46665 neurons) (Figure 6.2). Hyperparameters were optimized using the Talos Keras tuning module \cite{autonomio}. A Nvidia P6000 Quadro GPU card was used to train the model with the following hyperparameters: nonlinearity = relu, dropout rate = 0.2, optimization = AMSGrad, loss = mean squared error, batch size = 50, epochs = 100. Predicted output data were initially assessed using three metrics: root-mean squared deviation/error (RMSD/RMSE), mean squared error (MSE), and mean absolute percentage error (MAPE).

\textbf{Post-prediction processing} The predicted output data (flattened upper triangular data of pairwise interatomic distance matrices) were then transformed into symmetric pairwise interatomic distance matrices. This was then transformed into 3D coordinate data using the multi-dimensional scaling function from the scikit-learn python module and MDtraj \cite{mcgibbon2015mdtraj} to generate PDB structures (Figure \ref{NN_method2}). Chimera \cite{pettersen2004ucsf} was then used to add hydrogens to predicted PDB structures, which were then further processed to remove inaccurate chiral predictions. If more than half of C$\alpha$ centers were inaccurately predicted as R chiral centers (D-amino acids instead of L-aminio acids), this indicated the MDS portion predicted a reflection of the true coordinates. This was mitigated by reflecting all coordinates over the y-axis for predictions exhibiting an $\frac{R}{S}$ ratio greater than 1. If fewer than half of $\alpha$-carbons exhibited R chiral centers, reflecting coordinates was unnecessary. Instead, Chimera was used to switch side chain coordinates and the $\alpha$-hydrogen for all inaccurately predicted C$\alpha$ chiral centers. After checking for correct chirality for each residue, all conformations were further minimized for 75 conjugate gradient steps. 

\begin{figure}[H]
	\includegraphics[width=0.7\textwidth]{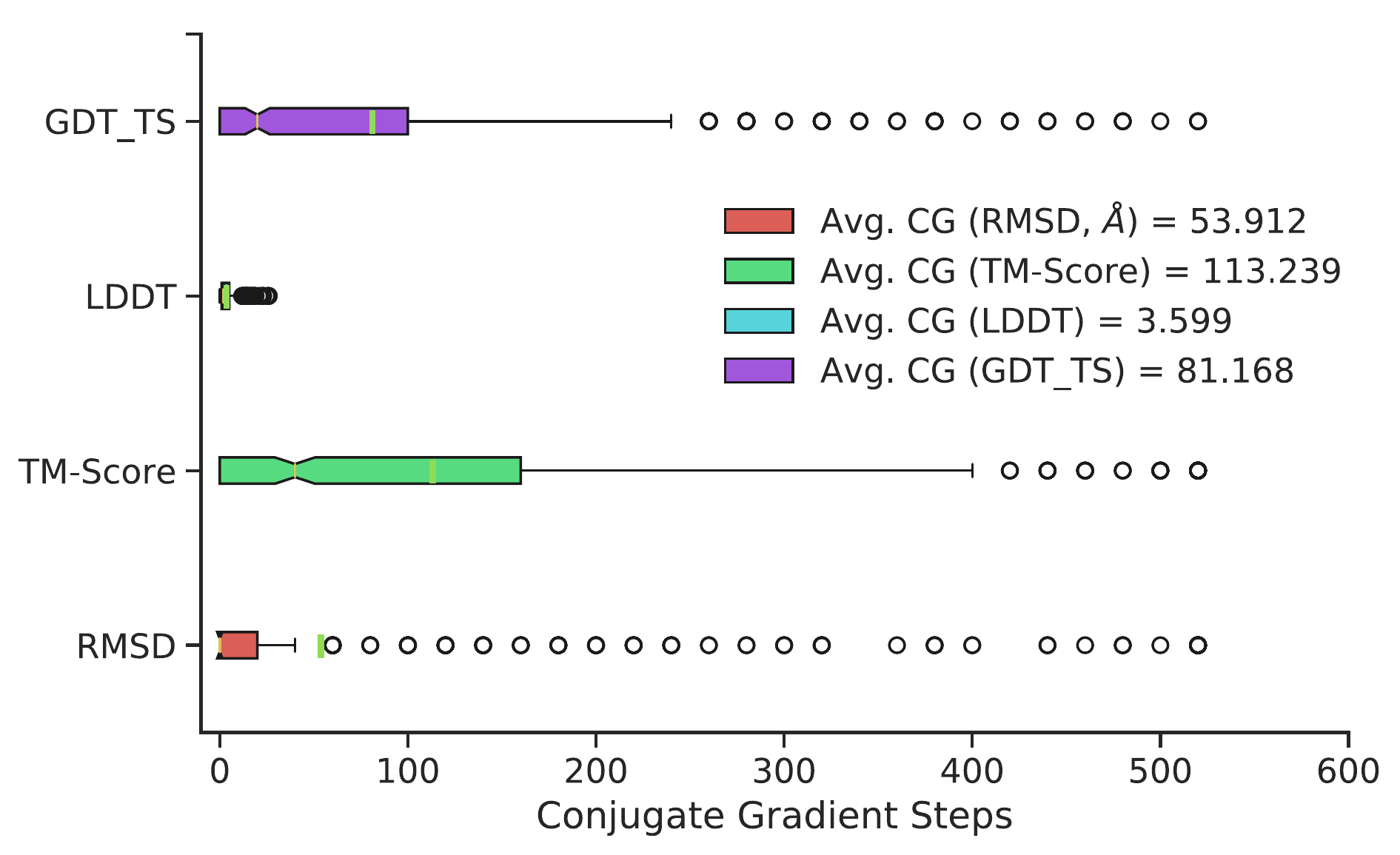}
	\caption[Distribution of the optimal number of CG steps for structure refinement, by metric]{Distribution of the optimal number of CG steps for structure refinement, by metric, with green vertical lines representing means, and notches representing medians.  Although exact optima vary by structure and metric, 50-100 steps are sufficient to provide good performance on most structures; extensive refinement beyond this point is rarely beneficial.}
	\label{NN_supp_metrics}
\end{figure}

The number of conjugate gradient steps was chosen by evaluating structures every subsequent 20 conjugate gradient steps for a cumulative 520 steps total. The maximum 520 conjugate gradient steps was chosen based on qualitative determination of average potential energy trends of all predicted conformations with increasing conjugate gradient minimization (Figure S\ref{NN_supp_metrics}). Three superposition-based metrics (RMSD, global distance test, total score (GDT\_TS), template modeling (TM) score) and one superposition-free metric (local distance difference test (LDDT)) were used to analyze any potential improvements in additional conjugate gradient steps between predicted 3D structure and the original, MD-generated 3D conformation. The RMSD metric analyzes all heavy atoms, TM score focuses primarily on C$\alpha$ atoms, and GDT\_TS also focuses primarily on backbone atoms. The LDDT score calculates a comparison using all-atom pairwise interatomic distances. Average values of 500 randomly chosen structures (RMSD, TM Scores, GDT\_TS, and LDDT) suggest a minimization range between 50-100 conjugate gradient steps. Thus, 75 steps was chosen as the total number of conjugate gradient steps to minimize all 11,926 predicted conformations. Overall, minimization yields minimial improvement relative to no minimization with respect to most metrics; however it is a  necessary step to remove steric clashes and slight stereochemical errors (Figure~\ref{NN_method2}, last panel). 

\section{Results} \label{sec_results}

\subsection{Multilayer perceptron (MLP) neural network reconstructs A\textbf{$\beta$} conformations with atomistic detail} 

\begin{figure}[H]
	\includegraphics[width=0.7\textwidth]{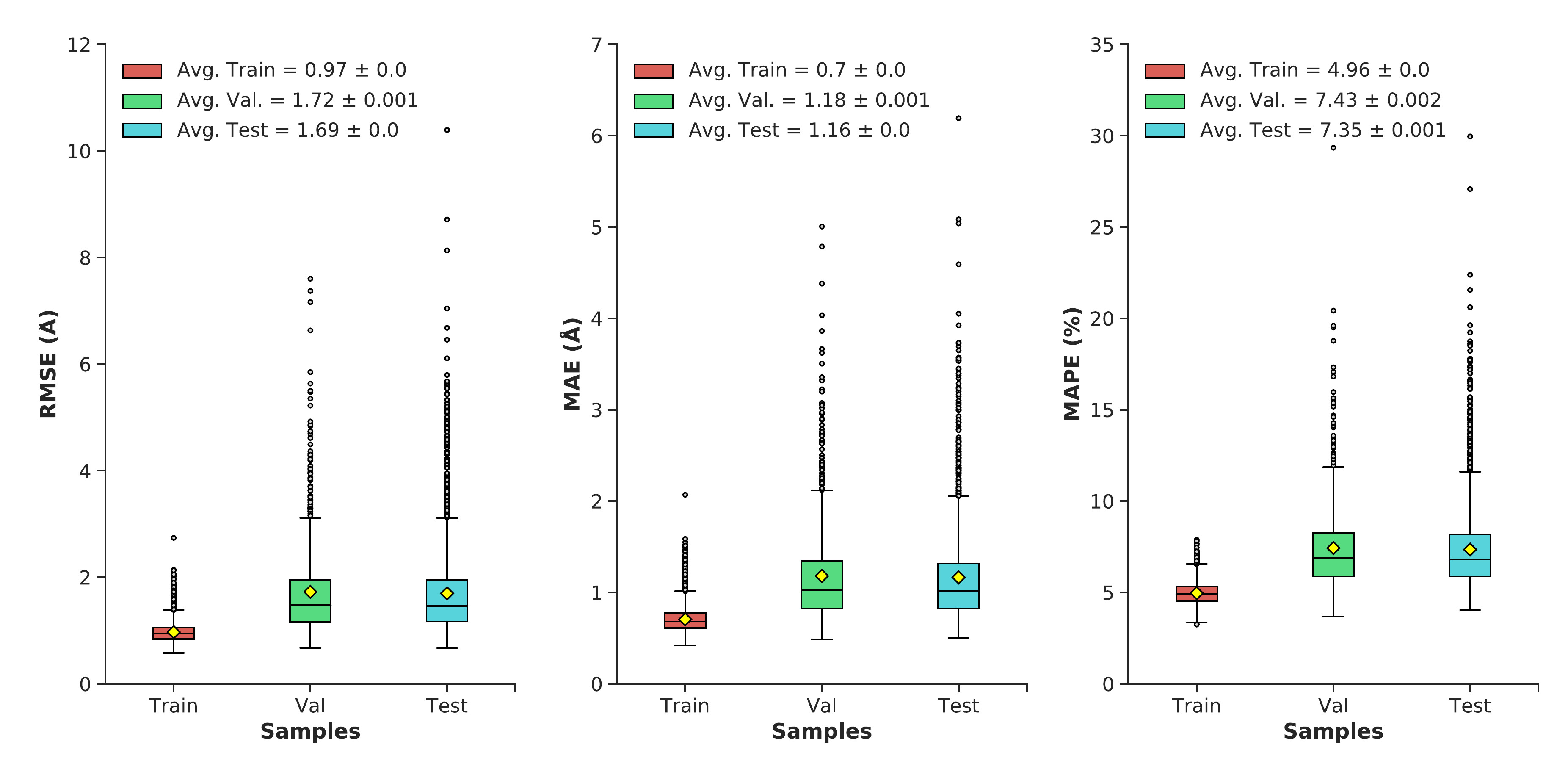}
	\caption[Boxplot distributions summarize the following metrics (RMSE, MAE, MAPE) for the train, validation, and test datasets.]{Boxplot distributions summarize the following metrics (RMSE, MAE, MAPE) for the train, validation, and test datasets: minimum, maximum, median, outliers (grey dots), average (yellow diamond) $\pm$ standard error, lower and upper quartiles.}
	\label{NN_boxplots}
\end{figure}

\begin{figure}[H]
	\includegraphics[width=0.7\textwidth]{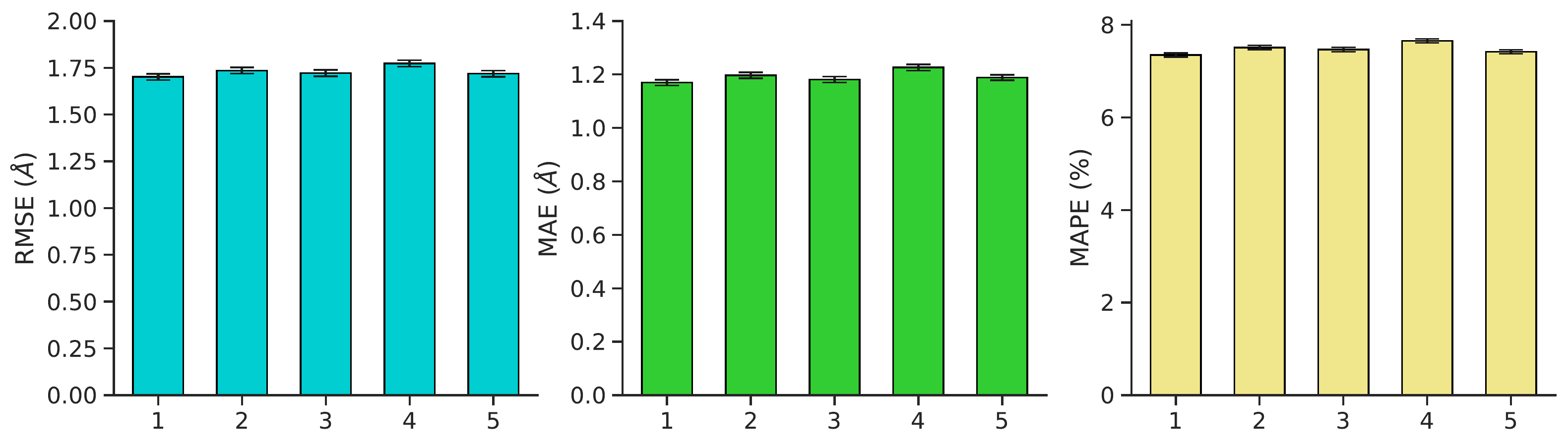}
	\caption[Validation performance under 5-fold cross-validation.]{Mean validation performance $\pm$ standard error on RMSE, MAE, and MAPE for each of five cross-validation splits. Performance is robust to choice of fold.}
	\label{NN_crossval}
\end{figure}

Pairwise interatomic distance (PID) predictions were made for all sets of data (train, validation, test). Predictions were evaluated against the ground-truth PIDs from the MD simulation using root-mean square error/deviation (RMSE/RMSD), mean absolute error (MAE), mean absolute percentage error (MAPE). The average metrics for the test set exhibit a favorable RMSE (1.7\AA), MAE (1.17\AA), and MAPE (7.35\%) (Figure \ref{NN_boxplots}). 5-fold cross-validation suggests bias was not arbitrarily introduced during the initial train-test split (Figure~\ref{NN_crossval}). Overall, average PID metrics for the validation and test set suggest the neural network was able to devise quality predictions. 

\begin{figure}[H]
	\centering
	\includegraphics[width=0.7\textwidth]{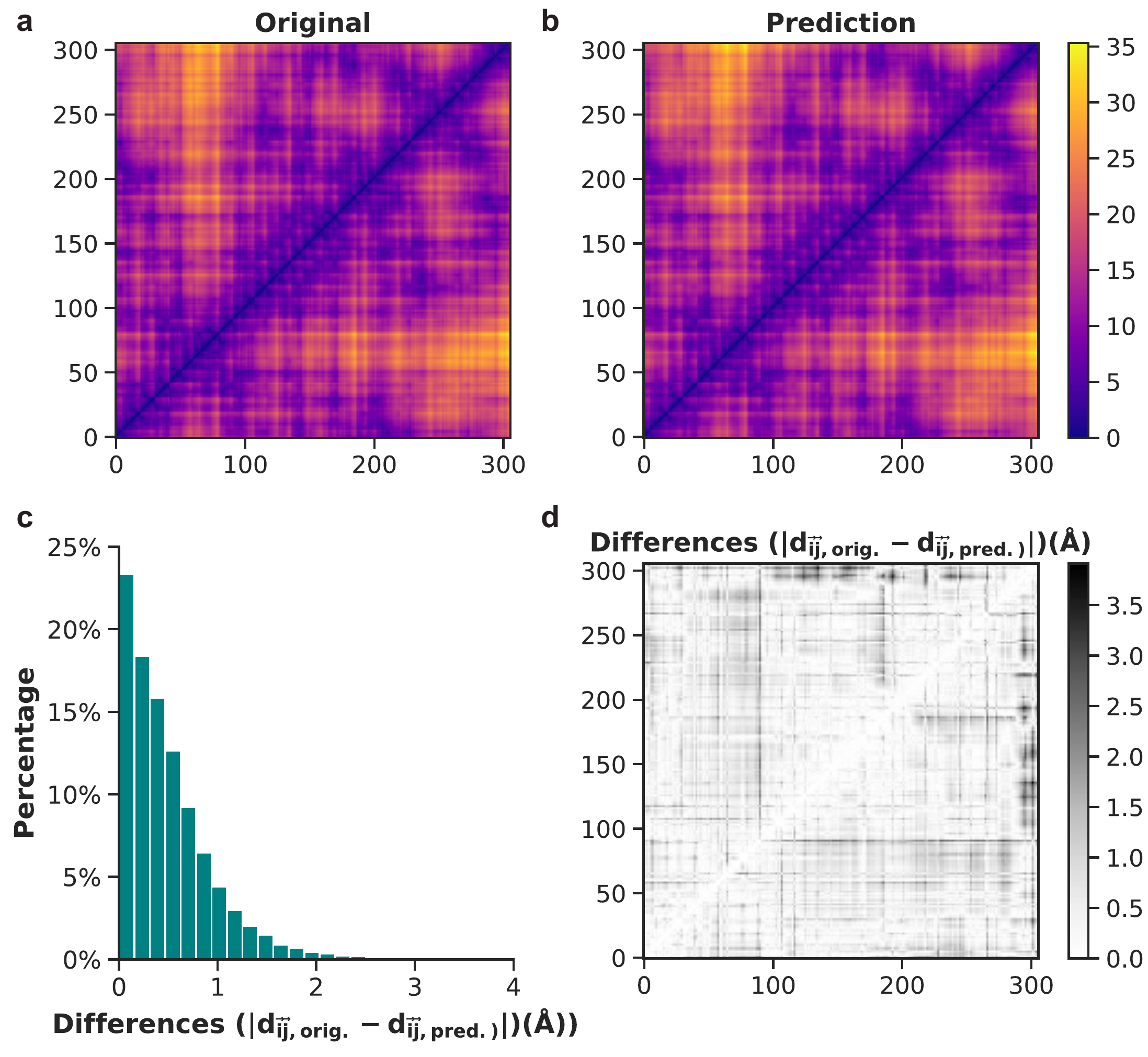}
	\caption[Comparison between original and predicted pairwise interatomic distances for frame 1133 (from the test set).]{\textbf{Comparison between original and predicted pairwise interatomic distances for frame 1133 (from the test set)}. \textbf{a.} Actual distances are shown for all heavy atoms. \textbf{b.} Heavy-atom predictions of all pairwise interatomic distance. \textbf{c.} Histogram of differences between original and predicted Euclidean distances. \textbf{d.} Binary plot displaying the absolute difference values between each actual and predicted distance for frame 1133.}
	\label{NN_distograms}
\end{figure}

To illustrate model performance, we show examples of both good and bad predictions from the test set, beginning with the positive example of frame 1133. Original and predicted pairwise interatomic distances for frame 1133 upon initial visualization, have highly comparable values (Figure \ref{NN_distograms}A-B). A grayscale depiction of absolute value differences between original and predicted PIDs reveals white and light grey data points, denoting mostly low values (Figure \ref{NN_distograms}D). A distribution of these data shows approximately 98\% of difference values are less than 2~\AA ~and 88\% are less than 1~\AA ~(Figure \ref{NN_distograms}C). Within the test set, this is an example of one of best-performing predictions made by the neural network model.  

\begin{figure}[H]
	\centering
	\includegraphics[width=0.7\textwidth]{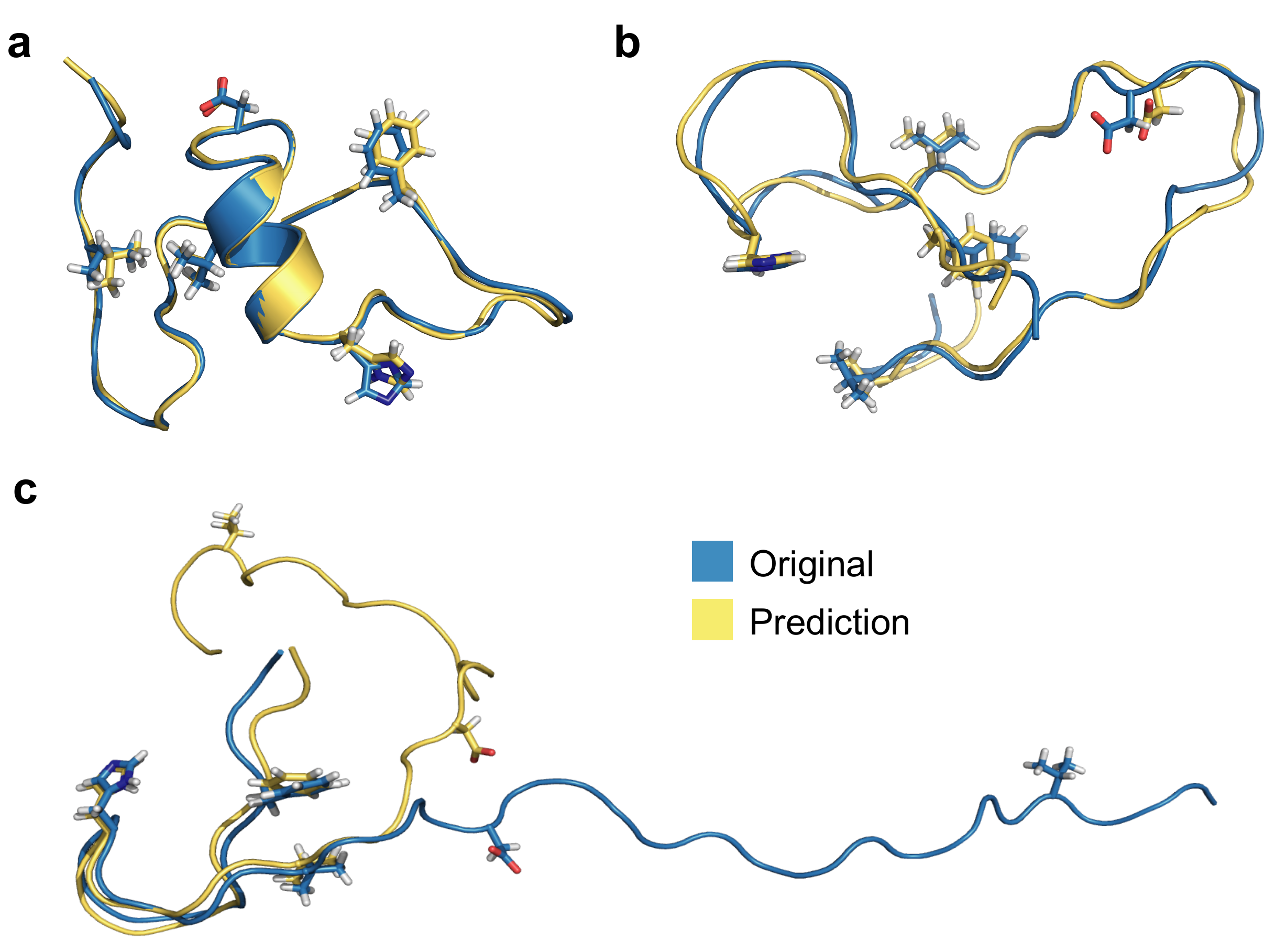}
	\caption[Alignment between original and predicted and processed 3D structures]{\textbf{Alignment between original and predicted and processed 3D structures} for (\textbf{a}) the best, (\textbf{b}) median, and (\textbf{c}) worst predictions based on RMSE values of PIDs. }
	\label{NN_3D_actual_predm}
\end{figure}

Using RMSEs of PIDs as a basis, we show processed 3D predictions of the lowest RMSE score representation (frame 1133, Figure \ref{NN_3D_actual_predm}A), the median representation (frame 7431, Figure \ref{NN_3D_actual_predm}B), and the highest RMSE score structure (frame 7560, Figure \ref{NN_3D_actual_predm}B). The best prediction with the lowest RMSE (0.67~\AA) exhibits more helical secondary structure compared to median and the worst predictions, which exhibit more random coil-like dynamics. RMSE of all heavy atoms for the median representation exhibits a fairly reasonable value of 1.46~\AA ~whereas the worst PID prediction has a RMSE of 10.4~\AA. Notably, the prediction for Figure \ref{NN_3D_actual_predm}C aligns reasonably well for the first 20 residues and the remaining residues are more poorly predicted by the neural network model. Because the PSN structure constrains folded regions more strongly than non-folded regions, it is not surprising the neural network model struggles to predict this specific overly extended conformation; however, we note that the prediction still preserves the qualitative aspects of the extended structure, and is quite accurate for the N-terminal region. The RMSEs according to 3D structure alignment between original and processed 3D structure and not on the basis of PIDs also contain similar values: best (0.77~\AA), median (2.13~\AA), and worst (12.01~\AA). These values are slightly higher compared to PID-based RMSEs, most likely due to introduced 3D alignment, whereas PIDs report RMSEs between all heavy atoms. 

\subsection{Generation of 3D structures and subsequent minimization}

When multidimensional scaling maps PIDs into 3D dimensional coordinates, it does so without regard to chirality. There are instances in which entire conformations are D- instead of L-amino acids, a correction that can be easily identified and fixed by reflecting coordinates across the y-axis. We also corrected conformations that contained only a few instances of D-amino acids, a result of the neural network predicting slightly incorrect side chain PIDs. These chirality checks followed by minimization are necessary, computationally inexpensive processing steps required  to transform PIDs into sterically reasonable 3D structures. Once corrections where fixed using Chimera, we then minimized all proteins for 75 conjugate gradient steps (a determination detailed in Methods), with a few conformations (23) requiring an additional 5 steps. 

\begin{figure}[H]
	\centering
	\includegraphics[width=0.7\textwidth]{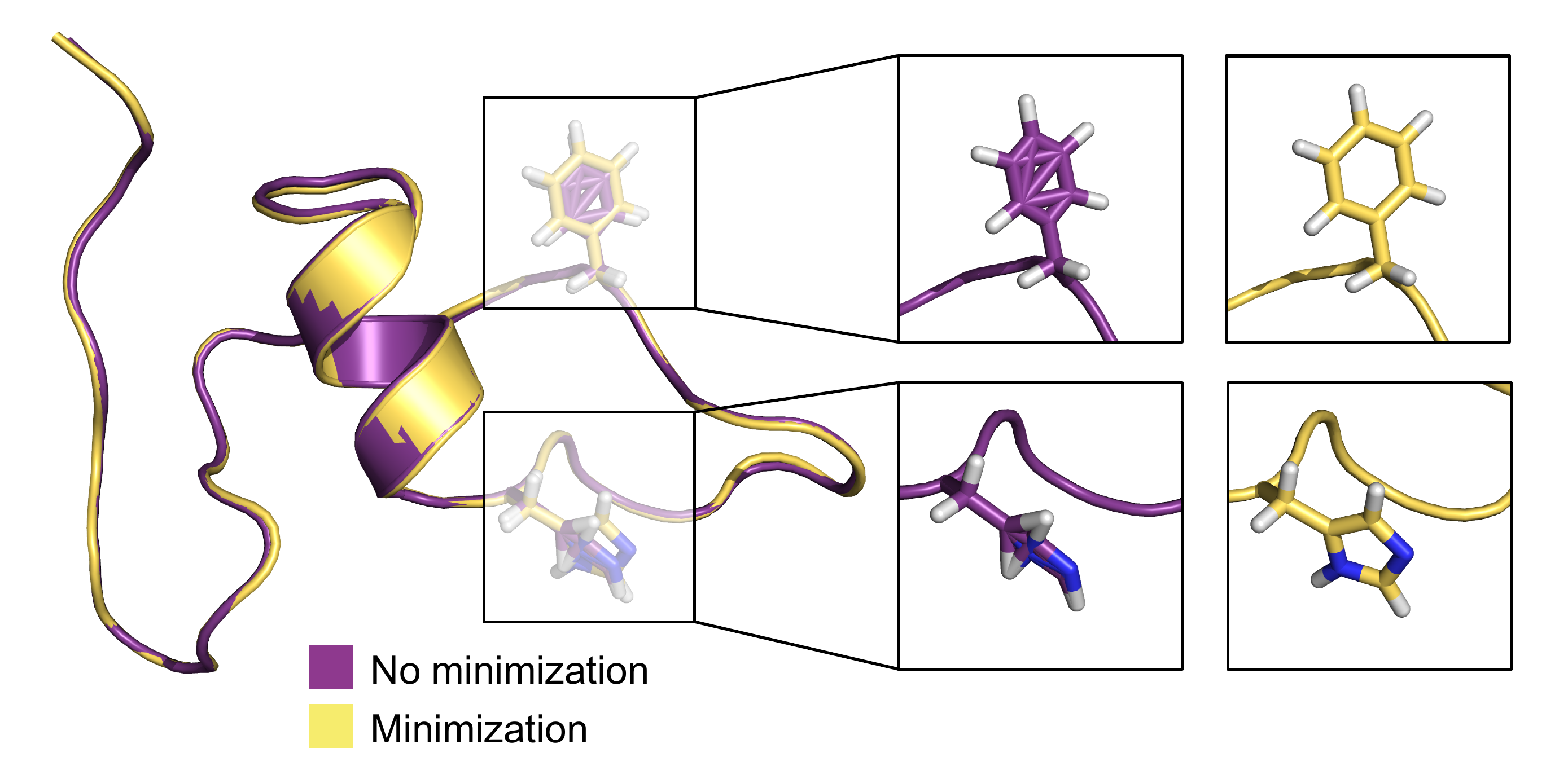}
	\caption[Comparison of pre- and post-minimized structures of the best prediction in the test set, frame 1133.]{Comparison of pre- and post-minimized structures of the best prediction in the test set, frame 1133.}
	\label{NN_min_nomin}
\end{figure}

Figure \ref{NN_min_nomin} depicts a pre- and post-minimization of the best predicted conformation (frame 1133) in the test set. Here we focus particularly on residues histidine 13 (His13) and phenylalanine 4 (Phe4). Both residues in the pre-minimized conformation are sterically incorrect and misplaced. Whereas in the post-minimized conformation, both residues have expected canonical sterics, devoid of incorrectly positioned atoms. When these optimization techniques (stereochemical corrections and minimization) are combined with the predictive power of the MLP neural network, this method yields highly effective predictive capabilities. 

\begin{figure}[H]
	\centering
	\includegraphics[width=0.7\textwidth]{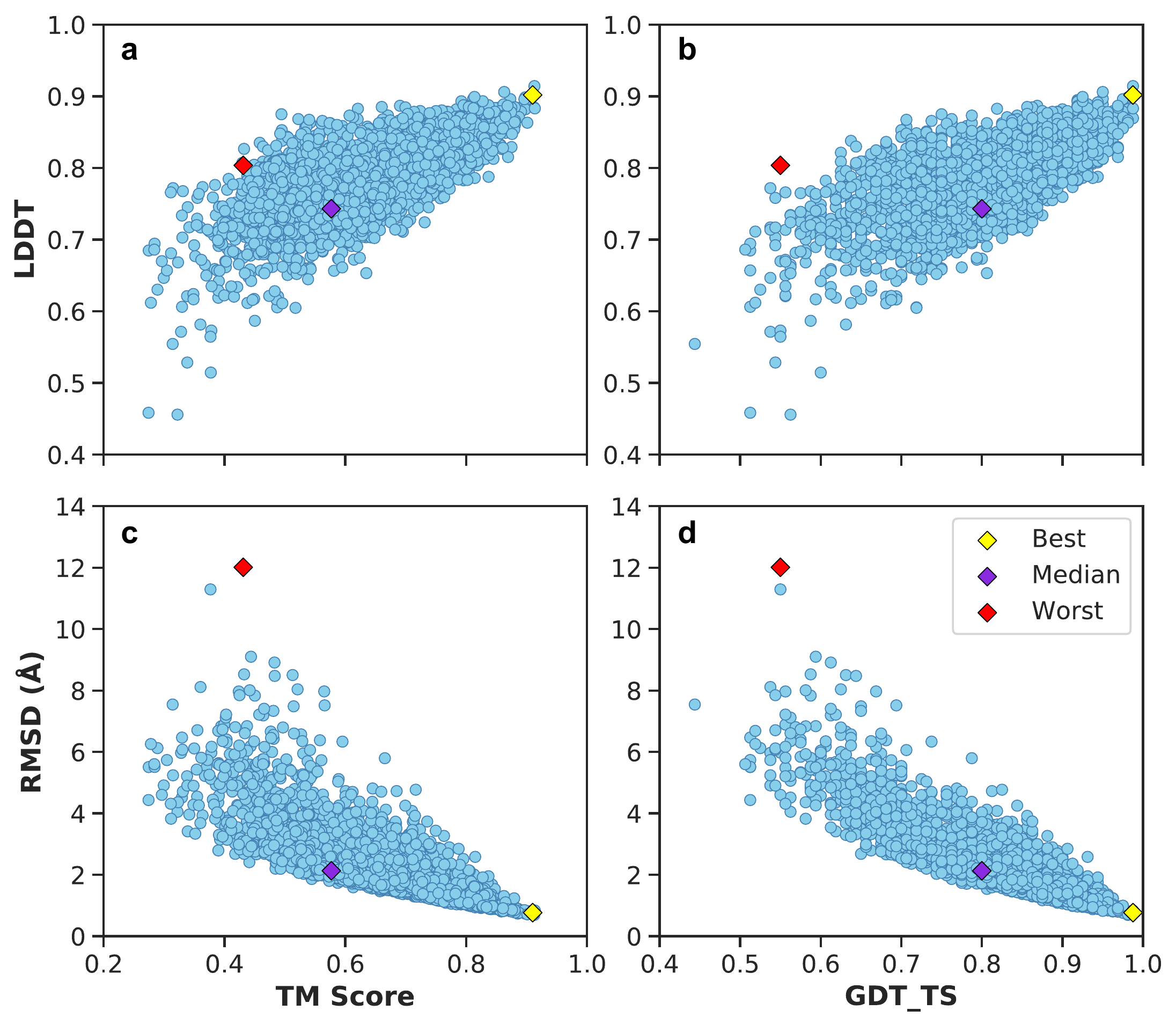}
	\caption[Juxtaposition of 3D structural metrics of the combined validation-test set: TM score, LDDT, GDT\_TS, and RMSD.]{\textbf{Juxtaposition of 3D structural metrics of the combined validation-test set: TM score, LDDT, GDT\_TS, and RMSD.} In addition, best, median, and worst predictions are shown based on PIDs. A) LDDT vs. TM score metrics of the validation-test set. B)  LDDT vs. GDT\_TS score metrics of the validation-test set. C) RMSD vs. TM score metrics of the validation-test set. D) RMSD vs. GTD\_TS score metrics of the validation-test set.}
	\label{NN_scores}
\end{figure}

After minimization, it was also imperative to compare 3D minimized predictions to their original MD simulation counterparts. Three superposition-based metrics (RMSD, TM score, GDT\_TS) and one superposition-free metric (local distance difference test (LDDT)) were utilized for this evaluation. The template modeling (TM) score measures the backbone similarity between a reference protein and target protein with a range from 0 (dissimilar) to 1 (identical) \cite{zhang2004scoring}. RMSD is a canonical protein comparison metric and here we parameterize it to compare all heavy atoms between native and predicted structures. LDDT utilizes pairwise interatomic distances in its methodology, focusing on local intramolecular interactions and the degree (range 0-1) of their retention in the target conformation in comparison to the native reference structure \cite{mariani2013lddt}. The global distance test, total score (GDT\_TS) is an improvement compared to RMSD designed to assess structures with the same sequence but different tertiary structure, with a higher score denoting better agreement (range 0-1) \cite{zemla2003lga}. All four metrics are commonly used during the biennial Critical Assessment of Structure Prediction (CASP) structure prediction and assessment competition \cite{kryshtafovych2019critical} and here we use these metrics to assess the predictive performance of the model. 

\begin{figure}[H]
	\centering
	\includegraphics[width=0.5\textwidth]{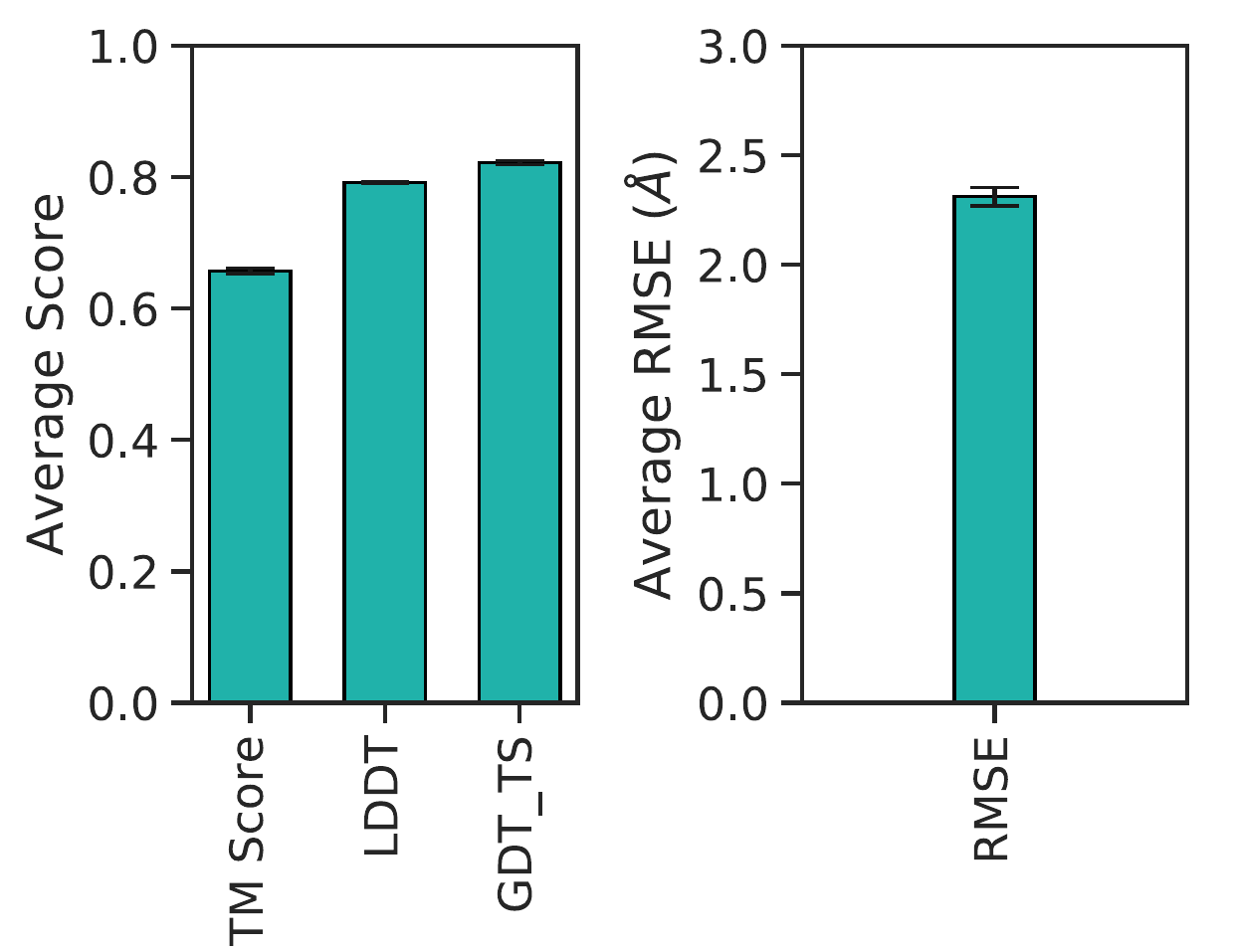}
	\caption[Barplot of average 3D accuracy metrics.]{Barplot of average 3D accuracy metrics and corresponding 95\% confidence intervals per score type.}
	\label{NN_scores_barplot}
\end{figure}

Figure \ref{NN_scores} illustrates these metrics for the combined validation-test set. There exists a positive correlation between LDDT vs. TM scores and GDT\_TS (Figure \ref{NN_scores}A-B). Between RMSDs vs. TM scores and GDT\_TS, predictions exhibit a negative correlation (Figure \ref{NN_scores}C-D).  Included are also the aforementioned best (yellow diamond), median (purple diamond), and worst (red diamond) PID predictions from Figure \ref{NN_3D_actual_predm}. Since their designation as best, median and worst were on the basis of RMSEs of PIDs and not 3D structure, it is interesting to observe the surprisingly high LDDT value of frame 7560 (the worst prediction). This suggests the neural network was able to preserve more local residue interactions despite struggling with larger more regional intramolecular interactions. TM scores exhibit values in the lower range of < 0.5, whereas most GDT\_TS and LDDT values occupy a range > 0.5, suggesting TM scores may not be as reliable of an assessment metric for A$\beta_{1-40}$. The average and 95\% confidence intervals suggest predicted 3D models are predicted relatively well considering the high GDT\_TS average and narrow 95\% confidence interval (Figure \ref{NN_scores_barplot}). The best and median test cases occupy expected 3D metrics (Figure \ref{NN_scores}). In combination with PID metrics (Figure \ref{NN_boxplots}), the 3D metrics demonstrate the model's ability to reasonably reconstruct the complex protein conformation of A$\beta_{1-40}$ from coarse contact adjacency matrices. 

\section{Discussion} \label{sec_disc}

In this work, we have implemented a custom MLP neural network model approach to reconstruct atom-level representations of A$\beta_{1-40}$ from residue-level PSNs. Although this particular neural upscaling model is specific to amyloid-$\beta$, the MLP neural network model can be retrained to other biomolecular systems from a variety of different sources (e.g. MD simulations, NMR ensembles, etc.), and thus can be generalizable and adaptable. For any given biomolecular coordinate structure, input (contact adjacency matrices) and output (PIDs) data for neural network retraining can be extracted.  More broadly, however, the success of the A$\beta_{1-40}$ model suggests that the general framework shown here can be applied to create similar models for other systems, or ultimately general-purpose upscaling models for broader classes of proteins.

Although previous reverse mapping methods (e.g. random placement, geometric-based, etc.) are able to reconstruct atomistic models, they do so typically from coarse grain force field models based on particle representations (e.g. MARTINI \cite{marrink2007martini}). The advantage of a MLP neural network is the ability to learn and fine-tune parameters specific to the system under investigation from minimal information (PSN adjacency matrices) in comparison to coarse grain force fields. The MLP neural network can thus familiarize itself with a specific target system of interest and coarse grain network simulations \cite{grazioli2019network} can be used to explore these biomolecules. 

In the literature, another class of neural networks, specifically variational autoencorders (VAE), have been used primarily on single small molecules and bulk-phase simulations as test cases for reverse mapping \cite{wang2019coarse}. This VAE methodology, although not tested on proteins, could possibly be adapted for such systems; however we are able to demonstrate successful backmapping with a non-variational MLP neural network architecture, indicating that variational structure is not essential. To better generalize our neural upscaling technique to protein systems of different sizes, convolutional neural network architectures similar to AlphaFold \cite{senior2020improved} could be also be incorporated and trained to predict regions (e.g. N x N residue regions).  With an ever-growing body of architectures whence to choose, there would seem to be considerable room for experimentation with alternative approaches.

Finally, we note that non-neural network methods can also be applied to the upscaling problem.  In preliminary experiments (not shown), we found that a kernelized ordinary least squares predictor \cite{scholkopf.smola:bk:2001} was able to obtain relatively good results (mean RMSD of approximately 2.4\AA, mean median ARE approximately 8\% on interatomic distances under 10-fold cross-validation).  Though the model was outperformed by the neural network architecture described here, and we did not therefore pursue it further, there may be situations in which non-neural network classes of predictors will prove useful.  This also  would seem to be a promising area for further investigation.

\section{Conclusions} \label{sec_conclusion}

Direct predictions of PID metrics demonstrate the predictive capabilities of the MLP neural network to reconstruct all-atom representations of proteins from binary contact adjacency matrices. Example conformations of the best, median and worst PID-based predictions in the test set illustrate the MLP performance. In the worst prediction (frame 7560), the RMSD between the N-terminal halves of the original vs. predicted is quite favorable (0.98 \AA). Chirality corrections and conjugate gradient minimization were vital post-prediction processing steps in generating stereochemically reasonable 3D structures. Three-dimensional accuracy metrics, in particular GDT\_TS -- the main assessment metric in the CASP competition -- suggests the neural network performed well given the average values and 95\% confidence intervals. In totality, we are able to illustrate the viability of the MLP neural network architecture in this transformation experiment. This work exemplifies neural network-based techniques capable of extracting useful, meaningful data from coarse grained models.

\vspace{6pt} 



\authorcontributions{Carter T. Butts performed MD simulation of A$\beta_{1-40}$, data processing, and PSN creation. Vy T. Duong created the NN architecture, trained the NN, processed and analyzed model performance. Gianmarc Grazioli contributed to study design.  Elizabeth Diessner contributed to data visualization.  Carter T. Butts and Rachel W. Martin conceived and designed the study.  All authors contributed to writing the paper.}

\funding{This research was funded by NSF awards DMS-1361425, IIS-1939237, and SES-1826589, and NASA award 80NSSC20K0620.}


\acknowledgments{We thank Ray Luo for co-advising of Vy Duong and use of computational resources. This research was also supported by computational resources supported by UCI Calit2 Think Tank.}

\conflictsofinterest{The authors declare no conflict of interest.} 


%

\end{paracol}
\reftitle{References}


\externalbibliography{yes}
\bibliography{bibliography}

\end{document}